\documentclass[10pt]{article}

\usepackage[utf8]{inputenc}
\usepackage{graphicx}
\graphicspath{{./figures/}}

\usepackage{amsmath,amssymb}
\usepackage{booktabs}
\usepackage{cite}
\usepackage{pgfplots}
\usepackage{siunitx}

\usepackage{csagh}

\makeatletter
\def\dump@received{\mbox{}}
\def\dump@revised{\mbox{}}
\def\dump@accepted{\mbox{}}
\makeatother

\pgfplotsset{compat=1.18}
\begin{document}
\begin{opening}

\title{Offline Ambient-Controlled Latent Diffusion: Architecture, Telemetry, and On-Device Evaluation}

\author[Callstack, Wroclaw, Poland; e-mail: lech.kalinowski@callstack.com; ORCID: 0000-0003-1143-3496]{Lech Kalinowski}
\author[AGH University of Krakow, Krakow, Poland; Callstack, Wroclaw, Poland; e-mail: amorys@agh.edu.pl; ORCID: 0000-0002-2137-8841]{Artur Morys-Magiera}
\author[Wroclaw University of Science and Technology, Wroclaw, Poland; Callstack, Wroclaw, Poland; e-mail: piotr.milkowski@callstack.com; ORCID: 0000-0001-5201-0364]{Piotr Mi{\l}kowski}

\begin{abstract}
Most mobile image-generation applications are thin clients over cloud services,
leaving outputs hard to audit. We present an Android latent-diffusion
application that runs entirely on-device and is driven by the ambient-light
sensor rather than a text prompt, keeping generation, telemetry, and storage
local. The contribution is not a new diffusion method but the surrounding
measurement workflow: each output is bound to the sensor reading, runtime path,
and seed that produced it, giving a per-artifact audit trail for offline
analysis. On a single Samsung foldable, one fixed capture of 373 artifacts
shows the controller's log-lux input positively associated with output
luminance (Pearson $r=0.532$, 95\% CI $[0.455, 0.601]$), confirming the ambient
dependency survives denoising and VAE decoding, while the latent UNet/VAE
pipeline runs at 552--1334\,ms mean latency across three quality tiers under the
Android Neural Networks API (NNAPI). The join logic and telemetry are released
for re-analysis.
\end{abstract}


\end{opening}

\section{Introduction}

Most mobile image-generation applications act as thin clients over cloud endpoints. This work studies what is required to move the full pipeline onto the phone --- model, sensing, and the instrumentation needed to study it --- and to drive it from the ambient-light sensor rather than a text prompt.

The central question is not merely whether the model runs locally. It is whether the resulting system is measurable: whether every output can be tied back to the sensor reading, runtime path, and seed that produced it, and whether that audit trail survives long enough to support offline analysis \cite{wohlin2024,madeyski2017,collberg2016}. This is a familiar concern in machine-learning systems engineering, where the surrounding software typically dominates operational complexity \cite{sculley2015,amershi2019}, but it has received less attention in on-device generative settings.

This paper describes the architecture, the instrumentation, and one fixed physical-device capture used to evaluate the system. The contribution is the measurement workflow --- the per-artifact telemetry, the artifact-to-trace join, and the offline analysis it enables --- rather than a new diffusion algorithm. We are explicit that the reported sensor-to-output correlation is not the contribution but a validation of it: because the controller is wired to mix ambient light into the latent update by construction (Eq.~\ref{eq:update}), the correlation serves to confirm that this documented dependency survives end-to-end to the recorded pixel statistics. The value of the work is that such a claim can be checked at all on an on-device generative pipeline, and re-checked from the released traces.

\section{Related Work}

\subsection{Experimental systems and reproducibility}

Experimental computer-systems research requires explicit measurement definitions and traceable analysis procedures \cite{wohlin2024}, and reproducibility remains a known weakness in the field \cite{madeyski2017,collberg2016}. Software-analytics work has argued that instrumentation should be treated as a first-class system component \cite{menzies2013,buse2012,fagerholm2017,ros2024}; we apply that view to a mobile generative pipeline.

\subsection{Systems engineering for AI-enabled applications}

Industrial studies of machine-learning products consistently report that the model is a small part of the operational surface \cite{sculley2015,amershi2019}. We study the same boundary in the specific case of on-device latent diffusion: model execution is evaluated alongside runtime selection, trace capture, and per-artifact analysis.

\subsection{Diffusion-based image generation and mobile runtime}

Denoising diffusion frameworks \cite{ho2020,song2021} and latent diffusion \cite{rombach2022} provide the algorithmic basis; latent-space denoising in particular makes mobile deployment realistic. Portable execution relies on runtime interoperability and hardware delegation, here via ONNX Runtime and the Android Neural Networks API \cite{onnxruntime,nnapi}. Quantization research \cite{han2016,jacob2018} offers a path to further efficiency, although the analyzed capture uses 32-bit floating-point models throughout.

On-device image synthesis is by now established rather than novel. Stable Diffusion was demonstrated running entirely on an Android phone, without a server, as early as 2023 \cite{qualcomm2023}, and fully local, prompt-driven generation on mobile neural accelerators has since become a small but real product category. We therefore do not claim on-device latent diffusion as a contribution: model execution on the phone is treated as available infrastructure. Two things distinguish the present system from that prior art. First, the control channel: generation is driven by the ambient-light sensor rather than by a text prompt, which the existing on-device applications do not address. Second, the instrumentation: every output is bound to the sensor reading, runtime path, seed, and pixel statistics that produced it, so the pipeline can be audited and replayed offline. The remainder of the paper develops that control-and-measurement workflow rather than the on-device execution it depends on.

\subsection{Ambient and context-aware interaction}

Context-aware and calm-computing work established environmental signals as control inputs for adaptive systems \cite{dey2001,weiser1996}. Using ambient light as the primary control channel for image synthesis is an extension of that principle; the novelty here lies in the surrounding execution and measurement system.

\section{System Architecture}

\subsection{Application design and scheduling}

\begin{figure}[!ht]
\centering
\includegraphics[width=\linewidth]{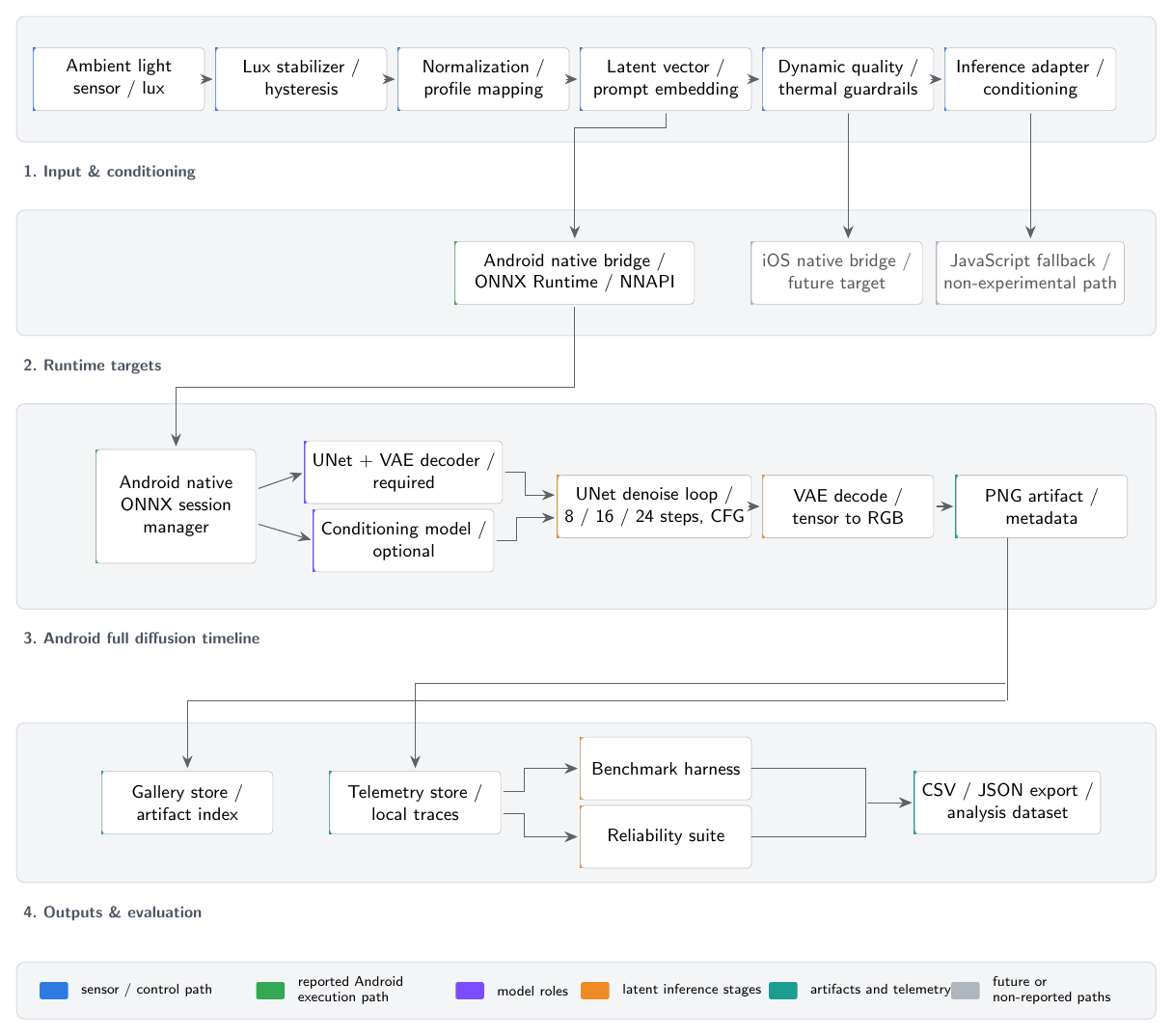}
\caption{Architecture overview: sensor control, on-device latent inference, artifact capture, telemetry, and reproducible export.}

\label{fig:architecture}
\end{figure}

The app has two screens: a picture-first ambient view and a diagnostics dashboard. Automatic generation runs only on the ambient view --- the dashboard pauses it so benchmark and reliability runs do not contaminate the trace. Triggers require both a temporal and a signal delta: at least 1\,s since the last generation, a lux delta of at least $\max(10, 0.006\,l_{prev})$, and a normalized-lux delta of at least 0.0025.

\subsection{Ambient conditioning pipeline}

Lux readings are stabilized by exponential smoothing and hysteresis, then mapped to a bounded control signal:

\begin{equation}
n(l)=\min\left(1,\frac{\log_{10}(1+l)}{\log_{10}(1+40000)}\right), \quad n(l)\in[0,1].
\label{eq:norm}
\end{equation}

The 40,000-lux anchor in the denominator corresponds approximately to direct outdoor sunlight; readings above it are clipped. Seven semantic profile anchors span the normalized range: Nocturne Bloom, Twilight Haze, Indoor Grain, Studio Drift, Overcast Veil, Soft Daylight, and Solar Echo.

\subsection{Runtime and inference orchestration}

At startup the app verifies local model artifacts and provisions role-specific sessions for conditioning, denoising, and decode. Session readiness flags and create/reuse/release counters are exposed through the telemetry layer so runtime behaviour can be inspected directly rather than inferred from latency.

\subsection{Latent conditioning rule}

The native denoising loop combines latent carry-over, ambient controls, conditioning features, and seeded perturbation. A compact schematic form is

\begin{equation}
\begin{aligned}
z_{t+1}={}&\lambda_t z_t + g_p(c_{profile}-0.5) + g_n(n(l)-0.5) \\
&+ g_r(\log_{10}(1+l)-0.5) + g_{\phi}(\phi(l)-0.5) \\
&+ g_f(\bar{f}-0.5) + b_{sem,t} + \eta_t + p_t
\end{aligned}
\label{eq:update}
\end{equation}

Here $\phi(l)$ denotes the raw-lux phase feature, $\bar{f}$ is the mean conditioning feature, $b_{sem,t}$ is the step-dependent semantic-bias term, $\eta_t$ is deterministic seed noise with step-dependent scale, and $p_t$ is the sinusoidal perturbation term. In the implemented controller, $g_p=0.11$, $g_n=0.08$, $g_r=0.06$, $g_{\phi}=0.04$, and $g_f=0.07$. The latent carry-over factor is dynamic, $\lambda_t=0.96-0.02\,b_{contrast,t}$, so Eq.~\ref{eq:update} should be read as the implemented controller around denoising rather than as a learned training objective.

\subsection{Measurement wiring}

Instrumentation is part of the architecture, not an afterthought \cite{buse2012,fagerholm2017}. Each generation event is persisted locally with latency, runtime path, seed, thermal status, guidance settings, ambient deltas, and rolling-latency context. Artifact metadata duplicates the most important fields so a row can be analyzed without joining. Benchmark and reliability harnesses write separate JSON reports.

An offline script joins artifacts to successful telemetry events and to pixel statistics extracted from PNGs, emits CSV datasets, and writes a separate audit file listing telemetry IDs it could not match. The unmatched IDs are preserved rather than dropped, on the principle that residual trace evidence is part of the result \cite{madeyski2017,collberg2016}.

\begin{table}[!ht]
\centering
\caption{Measurement layers and the fields they persist.}
\label{tab:wiring}
\footnotesize
\begin{tabular}{p{0.17\linewidth}p{0.36\linewidth}p{0.39\linewidth}}
\toprule
Layer & Persisted measures & Purpose \\
\midrule
Ambient control & Raw lux, normalized lux, profile, ambient delta, seed & Validate conditioning stability; quantify sensor-to-output coupling. \\
Runtime execution & Session flags and counters, runtime provider, step count, quality tier, stage latency, thermal status, guidance fields & Attribute latency and verify which pipeline executed. \\
Output artifact & Image size, output class, model variant, creation timestamp, pixel statistics & Enable deterministic replay and output-side analysis. \\
Experiment harness & Benchmark JSON, reliability JSON, join delta, source capture, unmatched telemetry IDs & Preserve audit trail and support repeatable analysis. \\
\bottomrule
\end{tabular}
\end{table}

\section{Experimental Methodology}

\subsection{Hardware and capture process}

Experiments ran on a single Samsung SM-F966B foldable, Android 16, 1080$\times$2520 display. Single-device evaluation limits external validity; we return to this in Section~\ref{sec:threats}. App-private artifacts and telemetry were exported with platform debugging tools and analyzed offline.

All quantitative results in this paper come from one fixed capture recorded on March 3, 2026, identified as \texttt{fold\_capture\_20260303\_134809}. It contains 373 generated artifacts, 377 telemetry events, 5 benchmark reports, and 3 reliability reports. Earlier mixed-variant captures exist in the repository for engineering history but are excluded here because they combine latent and procedural outputs.

\begin{table}[!ht]
\centering
\caption{Experimental protocol.}
\label{tab:protocol}
\footnotesize
\begin{tabular}{p{0.28\linewidth}p{0.62\linewidth}}
\toprule
Element & Description \\
\midrule
Device & Samsung SM-F966B, Android 16, 1080$\times$2520 display. \\
Capture date & March 3, 2026. \\
Dataset & 373 artifacts, 377 telemetry events. \\
Inclusion criterion & Artifact metadata reports model variant \texttt{ambient-latent-unet-vae-onnx-v1}. \\
Join rule & Successful telemetry matched to artifacts by seed equality and nearest timestamp within 180\,s. \\
Output statistics & PNG artifacts decoded for pixel-level luminance. \\
Latency filter & Robust subset excludes values below 100\,ms or above 10{,}000\,ms (7 of 373 rows). \\
Primary outcome & Correlation between log-transformed lux and mean image luminance. \\
\bottomrule
\end{tabular}
\end{table}

\subsection{Trace joining and reproducibility}

The join uses seed equality and nearest timestamp, with a 180-second tolerance ceiling. The ceiling was set defensively before the empirical delta distribution was known. In the reported capture, every accepted match fell within 58\,ms, so the join is effectively exact at the artifact level and unchanged under any reasonable tightening of the threshold. Four successful telemetry events did not match any artifact and are preserved in a separate audit file.

\subsection{Output synthesis}

Reported outputs come from the ONNX VAE decoder: decoded tensors are mapped to RGB and serialized as PNG. The decoder produces 320$\times$752 outputs in the configuration shipped with the capture; this is the VAE's native decode size for the deployed variant and is not a display crop. The implementation retains guarded fallback paths outside the scope of the reported runs.

\subsection{Model precision}

All active model artifacts in \texttt{assets/models/manifest.json} are 32-bit floating-point for the analyzed capture. No integer quantization or mixed precision is used in the numbers reported here; quantized variants are left for future work so their effect can be isolated from the ambient-control results.

\subsection{Evaluation metrics}

We report latency distribution, runtime-provider share for the latent-variant rows, latent-pipeline coverage, reliability pass rate (defined in Section~\ref{sec:results}), trace-join integrity, and the Pearson correlation between ambient measurements and image luminance. Latency aggregates use a 100--10{,}000\,ms inclusion window. Because the capture is one observational run, the reported correlations are descriptive of this dataset rather than causal estimates; Section~\ref{sec:threats} discusses what would be needed for causal claims.

\section{Results}
\label{sec:results}

\subsection{Execution scope}

\begin{table}[!ht]
\centering
\caption{Core experiment metrics.}
\label{tab:core}
\begin{tabular}{lr}
\toprule
Metric & Value \\
\midrule
Generated artifacts & 373 \\
Telemetry events & 377 \\
Benchmark reports & 5 \\
Reliability reports & 3 \\
Latent model variant rows & 373/373 \\
NNAPI runtime share among latent rows & 373/373 \\
Latency mean, robust subset & \SI{870.75}{ms} \\
Latency median, robust subset & \SI{770.50}{ms} \\
Latency p95, robust subset & \SI{1428.75}{ms} \\
Reliability pass rate & 15/15 successful \\
\bottomrule
\end{tabular}
\end{table}

Every artifact in the analyzed capture belongs to the target latent UNet/VAE variant and executed under NNAPI; the runtime-provider row in Table~\ref{tab:core} is not a separately enforced filter but the observed share for this capture, and therefore a useful sanity check that deployment did not silently fall back to CPU. The 15/15 reliability pass rate aggregates the three reliability reports, each of which records five independent startup-and-generate runs (fifteen runs in total). % AUTHOR: confirm the runs-per-report figure (3 reports x 5 runs = 15) matches the released JSON.
Each run exercises model loading, session creation, a single end-to-end inference, and artifact write, then checks that all four complete without error within a 12\,s budget; the test does not assess output quality.

Quality-tier counts were 173 preview, 87 balanced, and 113 high; robust mean latencies per tier were 552, 878, and 1334\,ms (Table~\ref{tab:latency}). Seven preview-tier rows exceeded 10{,}000\,ms and are excluded from the robust subset; no row in the released CSV fell below 100\,ms.

The cold and warm benchmark means coincide to within roughly 1\,ms (908.40 vs.\ 907.60\,ms). This is expected rather than anomalous: as described in the runtime-orchestration architecture, role-specific sessions are created and warmed once at application startup, before any measured generation, so even the ``cold'' benchmark loop runs against already-provisioned sessions. The label therefore distinguishes the first measured pass from later passes within a run, not an uninitialised pipeline from a warmed one; the absence of a warm-up gap is consistent with the session-reuse counters exposed by the telemetry layer.

\begin{table}[!ht]
\centering
\caption{Latency results, robust subset ($n=366$).}
\label{tab:latency}
\begin{tabular}{p{0.64\linewidth}r}
\toprule
Metric & Value \\
\midrule
Robust mean latency & \SI{870.75}{ms} \\
Robust median latency & \SI{770.50}{ms} \\
Robust p95 latency & \SI{1428.75}{ms} \\
Excluded outliers (of 373) & 7 \\
Preview mean / p95 & 552 / 851 ms \\
Balanced mean / p95 & 878 / 1155 ms \\
High mean / p95 & 1334 / 1467 ms \\
Benchmark cold mean & \SI{908.40}{ms} \\
Benchmark warm mean & \SI{907.60}{ms} \\
\bottomrule
\end{tabular}
\end{table}

\subsection{Ambient-to-output coupling}

\begin{figure}[!ht]
\centering
\begin{tikzpicture}
\begin{axis}[
width=4.3in,
height=2.6in,
xlabel={$\log_{10}(1+\mathrm{lux})$},
ylabel={Mean luminance},
ymin=0.20,ymax=0.75,xmin=1.0,xmax=5.2,
grid=both,grid style={line width=0.1pt, draw=gray!25},
tick label style={font=\footnotesize},
label style={font=\footnotesize}]
\addplot[only marks,mark=*,mark size=1.2pt,opacity=0.45,color=blue!70!black]
table[x=log10_lux,y=mean_luminance,col sep=comma]{csv/ambient1_20260303_experiment_dataset.csv};
\addplot[red,thick,domain=1.0:5.2,samples=2]{0.2050091326577043 + 0.09490822402426359*x};
\end{axis}
\end{tikzpicture}
\caption{Log-lux versus output luminance, $n=373$. Pearson $r=0.532$, 95\% CI $[0.455, 0.601]$.}
\label{fig:luxlum}
\end{figure}
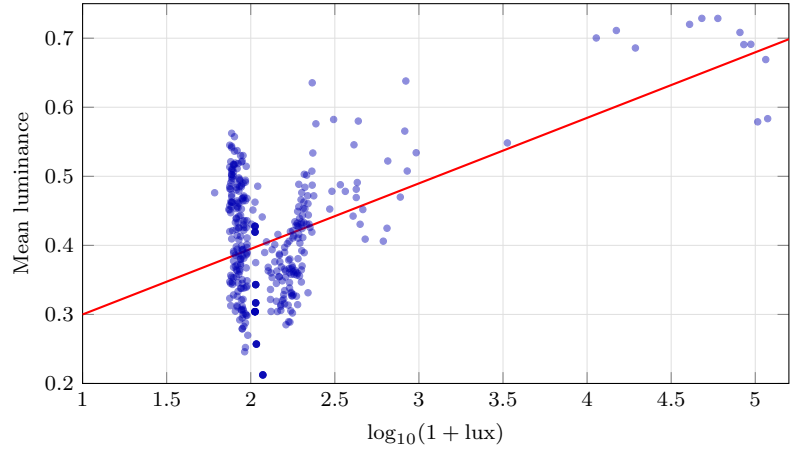

Figure~\ref{fig:luxlum} plots log-lux against mean output luminance. For the full latent capture ($n=373$), Pearson $r=0.532$ with 95\% CI $[0.455, 0.601]$. The result is a positive association between the controller's ambient input and the pixel-level luminance recorded from the decoded outputs.

We emphasize what this correlation does and does not show. Eq.~\ref{eq:update} explicitly mixes $\log(1+l)$ and $n(l)$ into the latent update through the ambient-control gains, so a non-zero relationship between log-lux and output luminance is built into the controller by design. The correlation in Figure~\ref{fig:luxlum} therefore functions as a wiring check: it confirms that the controller's documented dependency on ambient light survives through denoising and VAE decode to the pixel statistics. It is not, by itself, a causal estimate of visual brightening.

\subsection{Lux and latency}

Across the robust subset, log-lux and latency are inversely correlated (Pearson $r=-0.405$, $n=366$). Stratifying by quality tier shows that the association is heterogeneous rather than uniform: preview shows a weak negative relation ($r=-0.057$, $n=166$), balanced is effectively null ($r=0.023$, $n=87$), and high retains a stronger negative association ($r=-0.443$, $n=113$). The gap between the pooled coefficient and the within-tier coefficients is a textbook tier-composition (Simpson's-paradox) effect: the three tiers occupy different latency bands by construction (552, 878, and 1334\,ms means), and their rows are not uniformly distributed across the lighting range, so pooling mixes a large between-tier latency gap with comparatively weak within-tier lighting variation. The pooled coefficient should therefore be read as a property of the tier mix in this observational capture rather than as a lighting effect that holds within a fixed generation regime.

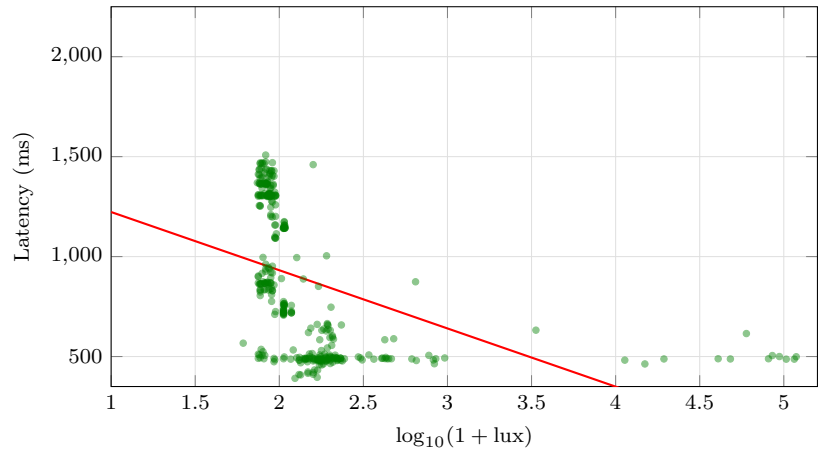
\begin{figure}[!ht]
\centering
\begin{tikzpicture}
\begin{axis}[
width=4.3in,
height=2.6in,
xlabel={$\log_{10}(1+\mathrm{lux})$},
ylabel={Latency (ms)},
ymin=350,ymax=2250,xmin=1.0,xmax=5.2,
restrict y to domain*=0:3000,
unbounded coords=discard,
grid=both,grid style={line width=0.1pt, draw=gray!25},
tick label style={font=\footnotesize},
label style={font=\footnotesize}]
\addplot[only marks,mark=*,mark size=1.2pt,opacity=0.45,color=green!50!black]
table[x=log10_lux,y=latency_ms,col sep=comma]{csv/ambient1_20260303_experiment_dataset.csv};
\addplot[red,thick,domain=1.0:5.2,samples=2]{1521.3967699850962 - 298.4127525600864*x};
\end{axis}
\end{tikzpicture}
\caption{Log-lux versus latency, robust subset ($n=366$). Pooled Pearson $r=-0.405$; within-tier correlations differ substantially (see text).}
\label{fig:luxlat}
\end{figure}

\begin{table}[!ht]
\centering
\caption{Correlation summary.}
\label{tab:correlations}
\begin{tabular}{p{0.43\linewidth}rr}
\toprule
Pair & $n$ & Pearson $r$ \\
\midrule
Log-lux vs.\ mean luminance, full capture & 373 & 0.532 \\
Log-lux vs.\ latency, pooled robust subset & 366 & $-0.405$ \\
Log-lux vs.\ latency, preview tier & 166 & $-0.057$ \\
Log-lux vs.\ latency, balanced tier & 87 & 0.023 \\
Log-lux vs.\ latency, high tier & 113 & $-0.443$ \\
\bottomrule
\end{tabular}
\end{table}

\begin{figure}[!ht]
\centering
\includegraphics[width=0.44\linewidth]{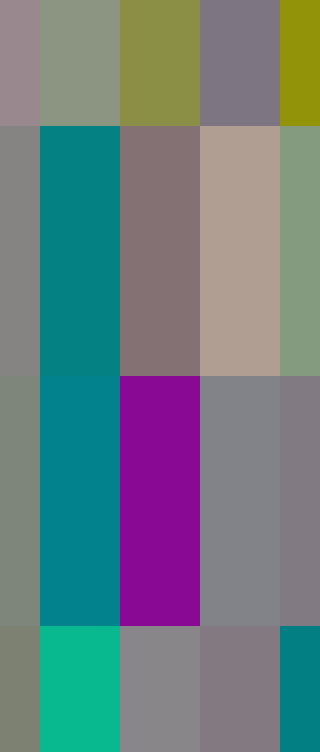}
\hfill
\includegraphics[width=0.44\linewidth]{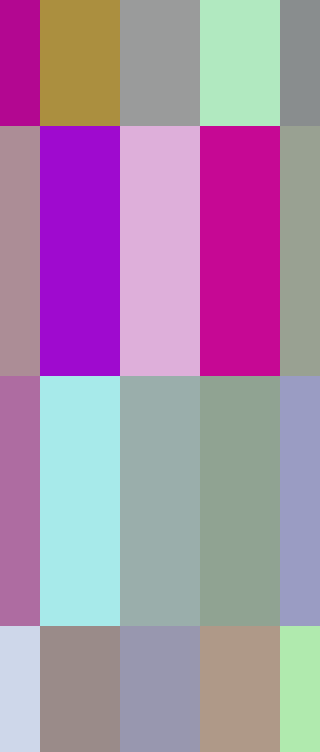}
\caption{Two VAE-decoded outputs at native 320$\times$752, one at 60 lux (left) and one at 118{,}519 lux (right). The right-hand reading lies above the 40{,}000-lux normalization anchor of Eq.~\ref{eq:norm}, so its normalized control value $n(l)$ saturates at 1.}
\label{fig:qual}
\end{figure}

\subsection{Trace integrity}

The artifact--telemetry join produced a complete artifact-level table and 4 unmatched successful telemetry events held in a separate audit file. The point is not that no rows were lost; it is that unmatched rows are listed by ID so they remain inspectable. The downstream CSV and the unmatched list together describe the full set of successful generations recorded during the capture.

\section{Discussion}

The headline result is that the ambient-control path is observable end-to-end: a sensor reading at the start of a generation event can be tied to the seed, the runtime path, the latency, and the pixel statistics of the resulting PNG. The positive Pearson correlation between log-lux and luminance confirms that the controller's wiring is intact through denoising and decode; it does not, on this observational capture, isolate ambient light as a cause of any broader aesthetic effect.

The latency picture is similarly structured. The tier ladder (preview 552\,ms, balanced 878\,ms, high 1334\,ms) is a reproducible result on this hardware. The pooled lux-latency coefficient is real for the robust subset, but it is not homogeneous across tiers: preview and balanced contribute little, whereas the high tier retains a stronger negative association. Reporting only the pooled number would therefore flatten an important part of the execution story.

\subsection{Threats to validity}
\label{sec:threats}

External validity is limited because the evaluation rests on one device. The framing throughout the paper is restricted to tested Samsung foldable hardware for this reason, and broader claims about Android-wide behaviour are out of scope. The capture is observational, so the reported correlations are descriptive rather than causal; a paired-scene, fixed-seed experiment with controlled coefficient variation would be required for causal claims. NNAPI metadata confirms the selected execution path but not which specific accelerator handled each operator; on Samsung devices this is typically a vendor-managed mix that is not directly observable from the app. Finally, the 180\,s join tolerance is much wider than the observed deltas in this capture, even though every accepted match fell within 58\,ms.

\section{Conclusion}

The paper described an Android latent-diffusion app driven by the ambient-light sensor and an instrumentation layer that makes its outputs auditable per artifact. On one Samsung foldable, the controller's log-lux input is positively associated with output luminance (Pearson $r=0.532$, $n=373$, 95\% CI $[0.455, 0.601]$), and the latent UNet/VAE pipeline runs at 552--1334\,ms mean latency under NNAPI across three quality tiers. The join logic and per-artifact telemetry are released so the analysis can be re-run; the natural follow-up is the paired-scene experiment that would convert this wiring-check result into a causal estimate.

\begin{acknowledgements}
This work received no external grant funding.
\end{acknowledgements}

\subsection*{Data and Code Availability}

The implementation, analysis scripts, and experiment data used in this paper are available in the project repository at \url{https://github.com/callstackincubator/hyper_ui/tree/codex/ambient-3}. The reported capture is identified as \texttt{fold\_capture\_20260303\_134809}; the manuscript's artifact-level dataset is provided under \texttt{paper/csv}.

\subsection*{Author Contributions}

Lech Kalinowski contributed 70\% of the work, including conceptualization, methodology, software, investigation, formal analysis, visualization, and preparation of the original manuscript. Artur Morys-Magiera contributed 15\% through methodological consultation, interpretation of results, and manuscript review and editing. Piotr Mi{\l}kowski contributed 15\% through technical consultation, validation of the experimental framing, and manuscript review and editing. All authors reviewed and approved the manuscript.

\subsection*{Financial Disclosure}

This work received no external grant funding. Hardware used in the experiments was acquired and operated independently; no party other than the authors' employer provided funding, hardware, or editorial input.

\subsection*{Conflict of Interest}

The authors declare no competing interests.

\subsection*{AI Assistance Disclosure}

Automated tools were used for language editing. All scientific content, implementation details, equations, and reported metrics were authored by the human authors and verified against the source repository and the captured experiment artifacts.

\bibliographystyle{cs-agh}
\bibliography{bibliography}

\end{document}